\documentclass[apj]{emulateapj}
\usepackage{apjfonts}
\begin{document}
\journalinfo{Accepted to ApJ}
\shorttitle{VLBI Observations of 6 GHz OH Masers in ON 1}
\shortauthors{Fish \& Sjouwerman}
\title{Global VLBI Observations of the 6.0~GH\lowercase{z} Hydroxyl Masers in Onsala~1}
\author{Vincent L.\ Fish\altaffilmark{1}
 \& Lor\'{a}nt O.\ Sjouwerman\altaffilmark{2}}

\altaffiltext{1}{MIT Haystack Observatory, Route 40, Westford, MA
  01886, USA, vfish@haystack.mit.edu}

\altaffiltext{2}{National Radio Astronomy Observatory, 1003 Lopezville
  Road, Socorro, NM 87801, USA, lsjouwer@nrao.edu}

\begin{abstract}
We present global VLBI observations of the first-excited state OH
masers in the massive star-forming region Onsala~1 (ON~1).  The 29
masers detected are nearly all from the 6035~MHz transition, and
nearly all are identifiable as Zeeman pair components.  The 6030 and
6035~MHz masers are coincident with previously published positions of
ground-state masers to within a few milliarcseconds, and the magnetic
fields deduced from Zeeman splitting are comparable.  The 6.0~GHz
masers in ON~1 are always found in close spatial association with
1665~MHz OH masers, in contrast to the situation in the massive
star-forming region W3(OH), suggesting that extreme high density OH
maser sites (excited-state masers with no accompanying ground-state
maser, as seen in W3(OH)) are absent from ON~1.  The large magnetic
field strength among the northern, blueshifted masers is confirmed.
The northern masers may trace an outflow or be associated with an
exciting source separate from the other masers, or the relative
velocities of the northern and southern masers may be indicative of
expansion and rotation.  High angular resolution observations of
nonmasing material will be required to understand the complex maser
distribution in ON~1.
\end{abstract}

\section{Introduction}

\object{Onsala~1} (ON~1) is a kinematically complex site of massive
star formation.  In the 1.6 and 6.0~GHz transitions of hydroxyl (OH),
masers are seen in two disjoint velocity ranges: $-2$ to
$+6$~km\,s$^{-1}$ in the north of the source and $+11$ to
$+16$~km\,s$^{-1}$ projected atop and to the south of the \ion{H}{2}
region \citep{argon00,fish05b,nammahachak06,fish07,fishreid07}.  A
4765~MHz excited-state maser at $+24.1$~km\,s$^{-1}$ was also reported
by \citet{gardner83} but never redetected.  The highly excited
13\,441~MHz transition shows similar velocity structure as the 1.6 and
6.0~GHz masers \citep{baudry02,fish05}, although an interferometric
map of the 13~GHz masers has not yet been published.  The systemic
velocity of the \ion{H}{2} region as measured by \citet{zheng85} in
H76$\alpha$ emission is approximately $+5$~km\,s$^{-1}$, which led the
authors to conclude that the redshifted masers seen atop the
\ion{H}{2} region are undergoing infall.  However, recent proper
motion measurements strongly suggest that the masers are tracing slow
($\sim 5$~km\,s$^{-1}$) expansion in ON~1 \citep{fishreid07}, similar
to what is seen in W3(OH)
\citep{bloemhof92,wright04,fishsjouwerman07}.

The ground-state OH masers in ON~1 have been observed with
connected-element interferometry
\citep{ho83,argon00,nammahachak06,green07} and with very long baseline
interferometry (VLBI) \citep{fish05b,fishreid07} on multiple occasions.
However, the only VLBI observations of the excited-state 6.0~GHz
masers were taken by \citet{desmurs98} with a three-station array.
Due to the poor image fidelity inherent with such a sparse array, they
were only able to detect 7 masers at 6035~MHz and 2 at 6030~MHz,
ranging in flux density from 0.5 to 7.1~Jy.  The large increase in the
number of telescopes with 6.0~GHz capability has since made
high-fidelity imaging of northern 6.0~GHz maser sources possible.

A prime motivation for observing ON~1 with a high sensitivity, high
angular resolution, high spectral resolution VLBI array derives from
the experience observing W3(OH), another nearby massive star forming
region.  In W3(OH), it was found that the 6030 and 6035~MHz OH masers
traced some areas where no ground-state masers have been observed and
highlighted portions of other areas where ground-state masers are
abundant, allowing a greater understanding of the large-scale
(Galactic cloud) structure delineated by molecules in the source
\citep{fishsjouwerman07}.  Of the massive star forming regions visible
from the north, ON~1 is one of the brightest 6.0~GHz OH maser sources,
with numerous masers detected in the \citet{baudry97} survey.  Thus,
in addition to being an interesting source per se, ON~1 is also an
interesting test case to determine to what degree OH masers in the
first excited state can provide information not available from the
ground-state masers alone.

\section{Observations and Data Reduction}

ON~1 was observed on 2008 Jun 15 with a global array consisting of two
National Radio Astronomy Observatory (NRAO) and seven European VLBI
Network (EVN) telescopes with $\lambda = 5$~cm capability (experiment
code GS029).  The participating stations were Effelsberg, the Green
Bank Telescope (GBT), a single Expanded Very Large Array (EVLA)
telescope, Hartebeesthoek, Jodrell Bank (Mark 2), Medicina, Onsala (85
ft), Toru\'{n}, and Westerbork (single telescope).

Both the 6030.747 and 6035.092~MHz main-line transitions of the first
excited state ($^2\Pi_{3/2}, J = 5/2$) of OH were observed in dual
circular polarization mode.  For each transition, a 500~kHz bandwidth
(24.8~km\,s$^{-1}$) was centered at an LSR velocity of
$+8$~km\,s$^{-1}$ in order to cover the full range of velocities of
maser emission previously detected in ON~1.  At correlation time, the
data were accumulated for 1~s intervals and channelized into 512
spectral channels, corresponding to a spacing of 0.049~km\,s$^{-1}$.
Both the parallel-hand and cross-hand polarization correlations were
obtained in order to reconstruct all Stokes quantities.

Data were reduced using the Astronomical Image Processing System
(AIPS).  The source 3C345 was used for instrumental delay and bandpass
calibration, and J2136$+$0041 was used for polarization calibration.
The polarization position angle of J2136$+$0041 was taken to be
$-162\fdg68$ from the average of C-band entries in the VLA/VLBA
Polarization Calibration
Database\footnote{\url{http://www.vla.nrao.edu/astro/calib/polar}}.
Visibility weights were modified by taking their fourth roots in order
to minimize the extent to which the overall calibration solutions
would be dominated by the Effelsberg-GBT baseline (and baselines to
these two highly sensitive telescopes).  Since OH masers usually
display substantial circular polarization, the ON~1 data were
self-calibrated using one circular polarization of a bright maser spot
in each transition, and the calibration solutions were applied to both
polarizations.

The source J2003+3034 was used to determine the absolute position of a
bright reference maser feature (listed in Table~\ref{maser-table}) at
each of the two frequencies.  Images were produced in all four Stokes
parameters as well as both right and left circular polarization (RCP
and LCP, corresponding to RR and LL correlations).  Maser spots were
identified in the RR and LL cubes where emission was above 6 times the
rms noise in each velocity channel.  For each maser feature, spectra
were obtained by integrating over a tight rectangular box encompassing
the feature emission.

\begin{deluxetable*}{lrrrrrrrl}
\tabletypesize{\scriptsize}
\tablecaption{Detected Masers at 6030 and 6035~MHz\label{maser-table}}
\tablehead{
  \colhead{} &
  \colhead{$v_\mathrm{LSR}$} &
  \colhead{R.A.} &
  \colhead{Decl.} &
  \colhead{Flux Density} &
  \colhead{$\Delta v$} &
  \colhead{Gradient} &
  \colhead{Gradient PA} &
  \colhead{Zeeman}
  \\
  \colhead{Pol.} &
  \colhead{(km\,s$^{-1}$)} &
  \colhead{Offset (mas)} &
  \colhead{Offset (mas)} &
  \colhead{(Jy)} &
  \colhead{(km\,s$^{-1}$)} &
  \colhead{(mas\,(km\,s$^{-1}$)$^{-1}$)} &
  \colhead{(\degr)} &
  \colhead{Pair}
}
\startdata
\tableline
\multicolumn{9}{c}{6035 MHz} \\
\tableline
R &   12.36 & $-$883.778 &    707.787  &  0.201 & 0.25 &\nodata&\nodata& A \\
L &   12.61 & $-$883.134 &    707.356  &  0.262 & 0.25 &\nodata&\nodata& a \\
R &   13.82 & $-$855.931 &    698.986  &  2.499 & 0.31 & 1.40 &    132 & B \\
L &   14.06 & $-$855.852 &    698.853  &  1.374 & 0.34 & 1.15 &    155 & b \\
R &   13.92 & $-$847.432 &    691.329  &  0.405 & 0.17 &\nodata&\nodata& C \\
R &   15.16 & $-$257.680 &    100.194  &  0.622 & 0.17 & 7.72 &  $-$70 & D \\
L &   15.42 & $-$258.373 &     99.586  &  0.475 & 0.18 &\nodata&\nodata& d \\
R &   15.18 & $-$250.056 &    102.927  &  1.276 & 0.26 & 1.48 &  $-$83 & E \\
L &   15.45 & $-$250.018 &    103.112  &  0.824 & 0.26 & 2.09 &  $-$41 & e \\
R &   13.98 & $-$227.783 &    102.493  &  0.195 & 0.20 &\nodata&\nodata& F \\
L &   14.28 & $-$227.264 &    101.794  &  0.241 & 0.19 &\nodata&\nodata& f \\
R &   14.78 & $-$202.760 &     87.994  &  0.229 & 0.24 &\nodata&\nodata& G \\
L &   15.07 & $-$203.000 &     87.858  &  0.218 & 0.24 &\nodata&\nodata& g \\
R &    5.51 & $-$128.465 &    994.596  &  2.229 & 0.26 & 1.16 &   $-$6 & H \\
L &    5.5\tablenotemark{b} & $-$128.492 &    995.516  &  0.227 &\nodata&\nodata&\nodata&(h)\\
R & $-$0.20 &  $-$49.760 &   1013.784  &  2.241 & 0.27 & 1.09 &  $-$54 & I \\
L &    0.48 &  $-$49.883 &   1013.685  &  1.702 & 0.27 & 0.89 &  $-$12 & i \\
R & $-$0.63 &  $-$47.799 &   1011.481  &  0.668 &\nodata&2.90 &    127 & J \\
R &   14.50 &      0.000 &      0.000  & 10.274 & 0.21 & 0.75 &    156 & K\tablenotemark{a}\\
L &   14.79  &      0.148 &   $-$0.515  &  0.320 &\nodata&\nodata&\nodata& k \\
R &   13.67 &     33.361 &    533.130  &  1.981 & 0.25 &12.52 &  $-$68 & L \\
L &   13.76 &     33.598 &    533.325  &  0.919 & 0.26 & 8.60 &  $-$72 & l \\
R &   14.40 &    148.685 &   $-$0.352  &  0.583 & 0.20 &\nodata&\nodata& M \\
L &   14.47 &    147.241 &      0.679  &  0.446 & 0.22 &\nodata&\nodata& m \\
R &   14.43 &    160.529 &   $-$6.392  &  1.147 & 0.21 &\nodata&\nodata& N \\
L &   14.50 &    160.754 &   $-$6.787  &  0.976 & 0.22 & 4.81 &    130 & n \\
\tableline
\multicolumn{9}{c}{6030 MHz} \\
\tableline
R &   13.79 & $-$855.753 &    699.107  &  3.244 & 0.28 & 1.11 &    169 & Z\tablenotemark{a} \\
L &   14.12 & $-$856.052 &    698.982  &  1.297 &\nodata&1.12 & $-$124 & z \\
L &   13.81 & $-$855.762 &    699.078  &  2.323 & 0.18 &\nodata&\nodata&(z)
\enddata
\tablecomments{Positions are offsets as measured from
  $20^\mathrm{h}10^\mathrm{m}09\fs0768, +31\degr 31' 34\farcs923$ (J2000).}
\tablenotetext{a}{Reference feature.}
\tablenotetext{b}{Velocity cannot be determined to an additional significant digit.  See text.}
\end{deluxetable*}

\section{Results}\label{results}

\begin{figure}
\resizebox{\hsize}{!}{\includegraphics{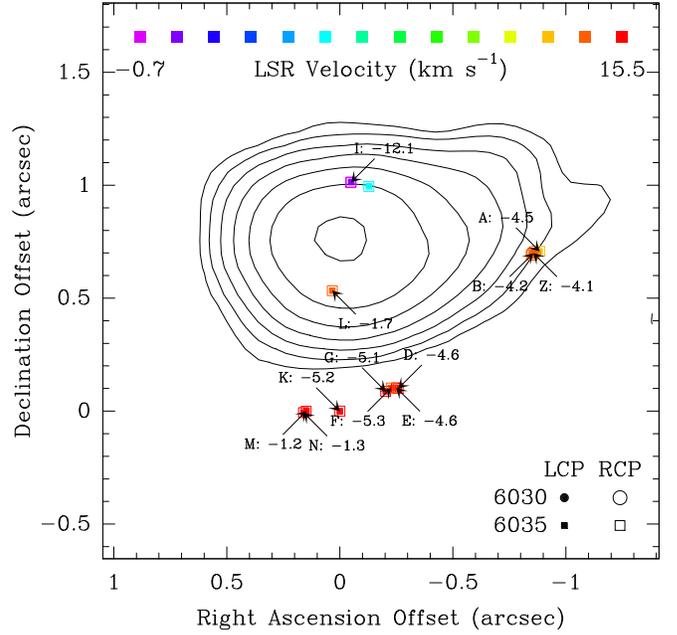}}
\caption{Map of 6.0~GHz masers in ON 1.  Squares and circles denote
  6035~MHz and 6030~MHz masers, respectively.  Maser lines identified
  in RCP are shown as open symbols, while LCP masers are shown as
  filled symbols.  Numbers indicate magnetic fields in milligauss as
  derived from Zeeman splitting, with letters denoting the associated
  features in Table~\ref{maser-table}.  Contours show 8.4~GHz
  continuum emission \citep{argon00}.  Offsets are measured from
  $20^h10^m09\fs0768, +31\degr 31' 34\farcs923$ (J2000).
\label{fig-map}
}
\end{figure}

\begin{figure}
\resizebox{\hsize}{!}{\includegraphics{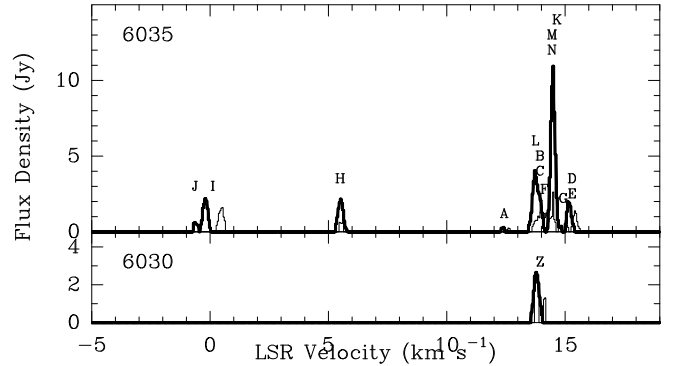}}
\caption{Total recovered maser emission in ON~1.  RCP is shown in bold
  and LCP in normal weight.  The same vertical scale is used for both
  panels.  Letters denote features in Table~\ref{maser-table}.
\label{fig-spectra}
}
\end{figure}

We detected 29 masers, of which the vast majority (26) are 6035~MHz
masers (Table~\ref{maser-table}).  The distribution of 6.0~GHz masers,
shown in Figure~\ref{fig-map}, confirms at higher resolution the
general distribution seen at lower resolution \citep{fish07}.  The
spectra of recovered emission (total flux density in spots identified
as masers) are shown in Figure~\ref{fig-spectra}.  The spectra are
very good qualitative matches to those obtained by \citet{green07}.

Neither 6035~MHz masers at $v_\mathrm{LSR} \approx +1$ or
$+2$~km\,s$^{-1}$ in the north of the source nor any low-velocity
6030~MHz masers at all are detected, despite their appearance in
previous studies of ON~1 \citep{baudry97,fish06,fish07}.  These masers
were weak (peak flux density 0.2~Jy, or approximately the flux density
of the weakest detected maser in this work) in the \citet{baudry97}
Effelsberg observations.  At least one such maser was seen above a
brightness level of 0.7~Jy\,beam$^{-1}$ in the EVLA observations of
\citet{fish07}, suggesting variability.  We note a possible RCP
6035~MHz feature at $(-164,+1049)$~mas, $v_\mathrm{LSR} =
+2.61$~km\,s$^{-1}$ in our data, but the spot is seen only in one
channel and only at $5.7\,\sigma$, which does not meet our
detectability criteria.

Nearly every 6030 and 6035~MHz maser feature is found with an
accompanying feature in the opposite polarization at a slightly
different velocity, presumably due to Zeeman splitting in the presence
of a magnetic field.  There are 12 or 13 readily identifiable Zeeman
pairs constituting 24 or 26 maser spots.  The high Zeeman pairing
efficiency (over 80\% of the detected 6~GHz masers can be paired) in
the 6~GHz transitions is consistent with the small Zeeman splitting
coefficients at 6030 and 6035~MHz as compared with the ground-state
transitions.  Observations of W3(OH) show a similar trend for
increased Zeeman pairing efficiency with decreased Zeeman splitting in
velocity units \citep{wright04b,fishsjouwerman07}.

\begin{figure}[t]
\resizebox{\hsize}{!}{\rotatebox{-90}{\includegraphics{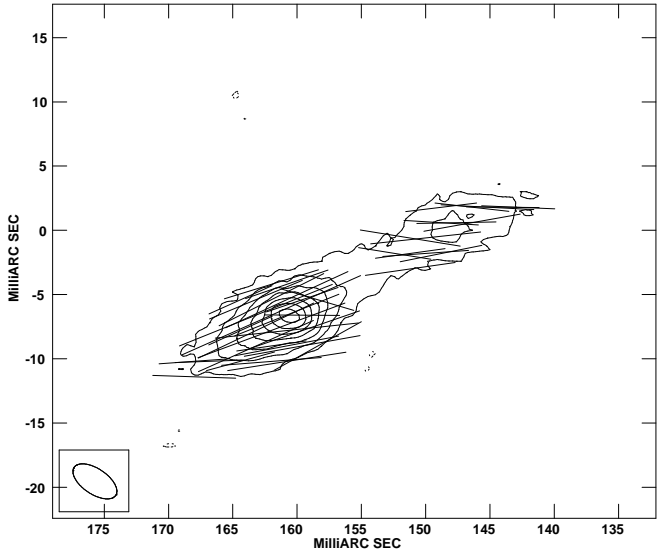}}}
\caption{Integrated Stokes~I emission and polarization of spots Mm and
  Nn in the velocity range 14.31 to 14.65~km\,s$^{-1}$.  Contours are
  shown as multiples of 4 times the rms noise.  Lines indicate the
  direction of linear polarization.
\label{fig-pcntr}
}
\end{figure}

We detect linear polarization at the $4\,\sigma$ level or greater in
several maser spots.  At 6030~MHz, spot Z shows a linear polarization
fraction ($\sqrt{Q^2+U^2}/I$) of 17\% with an electric vector
polarization angle of $-32\degr$ east of north in the channel of peak
emission, approximately consistent with \citet{green07}.  At 6035~MHz,
spot K is 22\% linearly polarized with a position angle of $52\degr$
and spot B is 17\% linearly polarized at $-19\degr$.  There is
evidence that the position angle of these spots may vary across the
emission line in frequency.  There is a ridge of weak emission in
eight consecutive velocity channels passing through spots Mm and Nn.
The linear polarization in spots Mm and Nn is not strong in any
individual velocity channel, but after integrating over all eight
velocity channels in which emission is detected in this region, linear
polarization is clearly detected and found to be aligned with the
ridge (Fig.~\ref{fig-pcntr}).

Positional gradients are computed via weighted least-squares fitting
of a straight line to a maser spot's centroid right ascension and
declination in each frequency channel in which it is detected.  The
positional gradients, given in Table \ref{maser-table}, indicate the
magnitude and position angle of the gradient (increasing in the
direction of positive velocity).  Maser components in Zeeman
associations have similar positional gradients.  It is possible that
the very large gradient of spot Ll results from the spatial blending
of two spots separated by less than the synthesized beamwidth.
Blending of nearby maser spots can make it difficult to determine
positional gradients when masers are not clearly isolated; for
instance, the positional gradient of spot n is roughly aligned with
the ridge of emission connecting it with spot m.  \citet{green07} note
a positional gradient aligned along a northwest-southeast direction in
three of five 6035~MHz Zeeman pairs, although it is not clear from
their text whether the ridge of spots Mm and Nn (which would have
appeared as a single spot at their resolution) is one of them.

\section{Discussion}\label{discussion}

\subsection{Zeeman Pairs}

The LCP and RCP components of 6~GHz Zeeman pairs are found to have a
very small spatial separation.  The separation between the centroid
positions of the peak channels of the LCP and RCP spots of a 6.0~GHz
Zeeman pair does not exceed 1~mas except for pair Mm, the weaker
component of the aforementioned ridge of emission in the southeast.  A
Zeeman component is identified for every 6.0~GHz maser whose emission
is not spatially blended with that of another, brighter spot.  The 2
spots labelled Hh may constitute a Zeeman pair, but the weakness of
the LCP feature prohibits measuring its velocity accurately enough to
identify the magnetic field strength and direction.

Only one Zeeman pair was identified at 6030~MHz.  It was found to be
spatially coincident with another Zeeman pair at 6035~MHz to within a
few tenths of a milliarcsecond.  The magnetic field and unshifted
velocity of the 6030 and 6035~MHz pairs agree to within 0.1~mG and
0.01~km\,s$^{-1}$, respectively.  Our magnetic field strengths are
broadly consistent with those of \citet{desmurs98}.  Differences on
the order of 1 to 2~mG can be explained by the lower spectral
resolution of the \citet{desmurs98} observations (0.11~km\,s$^{-1}$
channel separation = 1.4~mG at 6030 and 2.0~mG at 6035~MHz).  Our
magnetic field measurements are highly consistent with the MERLIN
observations of \citet{green07}, with differences well under 1~mG (no
greater than 0.2~mG in 5 of 6 Zeeman pairs).

\subsection{Multitransition Maser Associations\label{multi}}

\begin{figure}[t]
\resizebox{\hsize}{!}{\includegraphics{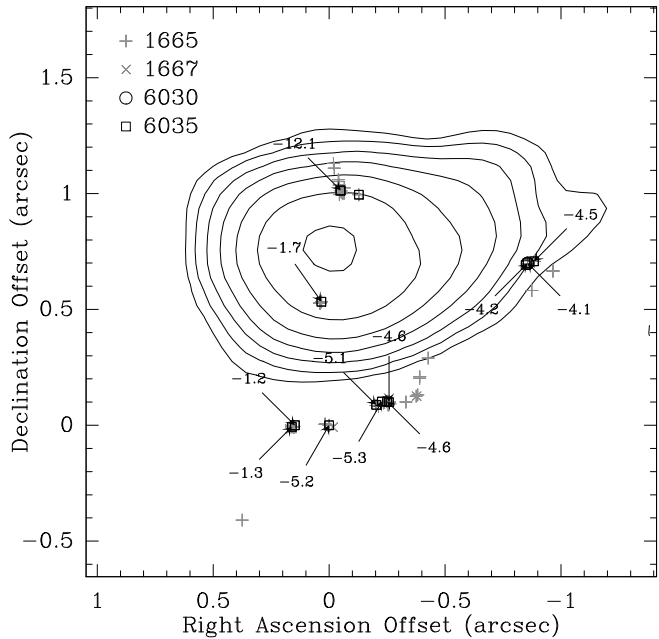}}
\caption{Relative locations of 6030/6035~MHz masers and ground-state
  masers \citep[in grey, from][]{fishreid07}.  Adopted alignment of
  the two data sets is discussed in \S~\ref{multi}.
\label{fig-multi}
}
\end{figure}

We find that nearly every 6.0~GHz maser detected in ON~1 is found in
close association with (i.e., less than 10~mas from) at least one
ground-state maser spot (Figure~\ref{fig-multi}) when the brightest
6035~MHz spot is aligned with the brightest 1665~MHz spot detected by
\citet{fishreid07}.  This result differs from the results of
\citet{green07}, who find from lower-resolution MERLIN data that the
closest 6.0/1.6~GHz separation involves a peak separation of 28~mas.
\citet{fishreid07} did not obtain absolute positions for the 1665 and
1667~MHz masers; nevertheless the separations between the 6.0~GHz and
nearest 1.6~GHz masers listed in Table~\ref{zeeman-table} are always
within a few milliarcseconds, and at least one associated ground-state
maser can be found for almost every 6.0~GHz maser found in this work.

\begin{deluxetable*}{lrrrrrrrrrr}
\tabletypesize{\tiny}
\tablecaption{Multitransition Zeeman Associations\label{zeeman-table}}
\tablehead{
  \colhead{6.0~GHz} &
  \colhead{6.0~GHz} &
  \colhead{6.0~GHz} &
  \colhead{6.0~GHz} &
  \colhead{1.6~GHz LCP} &
  \colhead{1.6~GHz LCP} &
  \colhead{1.6~GHz LCP} &
  \colhead{1.6~GHz RCP} &
  \colhead{1.6~GHz RCP} &
  \colhead{1.6~GHz RCP} &
  \colhead{1.6~GHz} \\
  \colhead{Zeeman} &
  \colhead{$B$} &
  \colhead{Separation} &
  \colhead{$v_\mathrm{LSR}^\mathrm{sys}$} &
  \colhead{R.A.} &
  \colhead{Decl.} &
  \colhead{$v_\mathrm{LSR}$} &
  \colhead{R.A.} &
  \colhead{Decl.} &
  \colhead{$v_\mathrm{LSR}$} &
  \colhead{$B$} \\
  \colhead{Group} &
  \colhead{(mG)} &
  \colhead{(mas)} &
  \colhead{(km\,s$^{-1}$)\tablenotemark{a}} &
  \colhead{Off.\ (mas)} &
  \colhead{Off.\ (mas)} &
  \colhead{(km\,s$^{-1}$)} &
  \colhead{Off.\ (mas)} &
  \colhead{Off.\ (mas)} &
  \colhead{(km\,s$^{-1}$)} &
  \colhead{(mG)\tablenotemark{b}} \\
}
\startdata
A &  $-$4.5 & 0.77 & 12.48 & $-$884.60 &    707.53 &   13.75 &  \nodata  &  \nodata  & \nodata &  $-$4.3 \\
B &  $-$4.2 & 0.15 & 13.94 & $-$855.70 &    697.28 &   15.15 &  \nodata  &  \nodata  & \nodata &  $-$4.1 \\  
Z\tablenotemark{c} &  $-$4.1 & 0.32 & 13.95 & & & & & & & \\
D &  $-$4.6 & 0.92 & 15.29 & $-$258.85 &    100.60 &   16.73 & $-$259.45 &    100.83 &   13.92 &  $-$4.8 \\
E &  $-$4.6 & 0.19 & 15.32 & $-$249.24 &    102.42 &   16.73 & $-$249.72 &    103.36 &   13.68 &  $-$5.2 \\
F &  $-$5.3 & 0.87 & 14.13 & $-$233.76 &     97.58 &   16.35 &  \nodata  &  \nodata  & \nodata &  $-$7.5 \\
G &  $-$5.1 & 0.28 & 14.92 & $-$201.74 &     85.70 &   15.88 &  \nodata  &  \nodata  & \nodata &  $-$3.2 \\
H &\nodata &  0.92\tablenotemark{d}&  5.5\phantom{0} &  \nodata  &  \nodata  & \nodata & $-$128.22 &    994.63 &    5.95 &  $+$1.7 \\
I & $-$12.1 & 0.10 &  0.14 &  $-$50.68 &   1012.23 &    3.90 &  \nodata  &  \nodata  & \nodata & $-$12.8 \\
K &  $-$5.2 & 0.54 & 14.64 &  \nodata  &  \nodata  & \nodata &   $-$0.02 &   $-$0.02 &   12.82 &  $-$6.1\tablenotemark{e} \\
L &  $-$1.7 & 0.31 & 13.72 &     40.34 &    526.99 &   13.88 &     39.19 &    527.41 &   13.21 &  $-$1.1 \\
N\tablenotemark{f} &  $-$1.3 & 0.45 & 14.47 &    154.58 &   $-$6.23 &   15.05 &    152.31 &   $-$4.82 &   14.02 &  $-$1.7\tablenotemark{g} \\
  &         &      &       &    161.92\tablenotemark{h} &  $-$13.37 &   14.64 &    160.68\tablenotemark{g} &  $-$13.49 &   14.29 &  $-$1.0 
\enddata
\tablecomments{The columns 6.0~GHz and 1.6~GHz refer to the 6035~MHz and 1665~MHz transitions,
respectively, except as noted in the table.  The 1.6~GHz maser information is taken from Table~2 in
\citet{fishreid07}.}
\tablenotetext{a}{Systemic velocity (corrected for Zeeman splitting).}
\tablenotetext{b}{When two 1.6~GHz spots are given, magnetic field is as derived from 1.6~GHz Zeeman pair.
When only one 1.6~GHz spot is given, magnetic field is inferred from multitransition overlap.}
\tablenotetext{c}{6030~MHz}
\tablenotetext{d}{Interpreting the RCP and LCP features as Zeeman components.}
\tablenotetext{e}{Paired with a distant LCP feature to give a $-2.5$~mG field in \citet{fishreid07}.}
\tablenotetext{f}{Spatially blended with Zeeman group M, which implies an almost identical magnetic field strength and systemic velocity.}
\tablenotetext{g}{Alternatively, the 6035~MHz spots may be grouped with (162.13,$-$7.20)~mas LCP 1665~MHz to give a $-1.0$~mG field.}
\tablenotetext{h}{1667~MHz}
\end{deluxetable*}

The close spatial associations allow for multitransition Zeeman
associations to be identified.  The Zeeman effect shifts LCP and RCP
components in frequency by the magnetic field strength multiplied by a
constant splitting factor that is different for each transition.
Under the assumption that multitransition associations are produced by
the same cloudlets of material, the magnetic field strength and
systemic velocity can be derived so long as any two Zeeman components
can be identified from amongst all transitions seen in spatial
overlap.  Table~\ref{zeeman-table} lists the magnetic fields derived
from Zeeman associations including both 6.0~GHz and 1.6~GHz masers.
There are four Zeeman associations in which a complete Zeeman pair is
found at both 6035~MHz and 1665~MHz (and one additional pair at
1667~MHz).  The magnetic fields obtained in these transitions are
consistent with each other in all cases to 0.6~mG or better.  The
excellent agreement between magnetic field strengths obtained from
1.6~GHz Zeeman pairs and 6.0~GHz Zeeman pairs provides strong evidence
supporting the alignment between the reference frames of these
observations and the \citet{fishreid07} VLBA observations.

Furthermore, Zeeman associations consisting of a pair of LCP and RCP
components at 6030~MHz and a single unpaired component at 1665~MHz
show that the implied magnetic field at 1665~MHz (assuming that the
systemic velocity is given by the average of the 6035~MHz LCP and RCP
velocities) is consistent with the 6035~MHz value.  In the six such
associations, the maximum difference between the 6035~MHz and implied
1665~MHz magnetic fields is 2.2~mG, with four associations showing a
difference of less than 1~mG.  We refer to these instances as
``multitransition Zeeman associations'' and argue that the above
results support the validity of using these to derive magnetic fields.

Of note is a potential $+1.7$~mG magnetic field implied by the overlap
of a 6035~MHz RCP spot (H) with a 1665~MHz RCP spot at a different
velocity.  There is also a very weak 6035~MHz LCP spot (h) at this
position whose velocity cannot be determined to sufficient accuracy to
identify the sign of the magnetic field from the 6035~MHz Zeeman pair
alone.

There does not appear to be a significant correlation between the
brighter polarization at 6035~MHz and that in a nearby ground-state
transition, usually 1665~MHz.  When the flux ratio between Zeeman
components at 6035~MHz is close to unity ($\lesssim 2.5$), there is no
correlation at all.  Only two 6035~MHz Zeeman pairs have a component
flux ratio greater than 2.5: Hh and Kk.  In both instances, the
brighter Zeeman component at 6035~MHz is RCP, and the RCP feature at
1665~MHz is detected, but the LCP feature is not \citep{fishreid07}.
In the one case where a 6030 and 6035~MHz spot overlap, the RCP
component is brighter than the LCP component in both transitions, and
the RCP/LCP ratio is greater for the 6030~MHz transition.  (However,
emission at 1665~MHz which is spatially coincident with these Zeeman
pairs is seen only in LCP, not RCP.)  There is also one instance where
the 1665, 1667, and 6035~MHz masers overlap, and the Zeeman component
flux ratios decrease in that order.  This is consistent with the flux
ratios varying monotonically with Zeeman splitting ratio
\citep{wright04b}.  There is also a similar trend for 6.0~GHz Zeeman
pairs to have smaller \emph{separation} between LCP and RCP components
than is seen for coincident ground-state transitions \citep[an effect
  also noted in W3(OH) by][]{fishsjouwerman07} in W3(OH).  These
effects may be related if, for instance, excitation conditions (such
as velocity coherence) change slightly over the physical extent of a
Zeeman association.

\begin{figure}
\resizebox{\hsize}{!}{\includegraphics{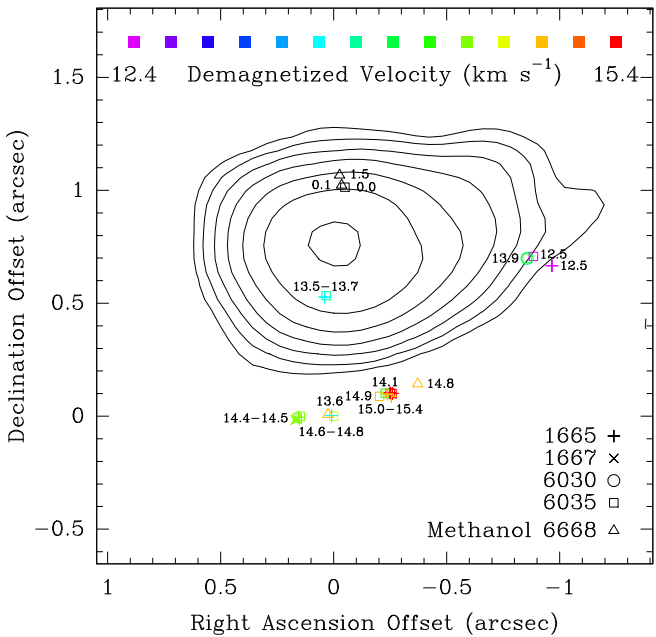}}
\caption{LSR velocities corrected for Zeeman splitting.  Color
  indicates the central (unshifted) velocity for Zeeman pairs in each
  transition (also indicated with numbers), with northern masers
  ($v_\mathrm{LSR} \ll 12.4$~km\,s$^{-1}$) in black.  Multitransition
  Zeeman associations are not shown.  Methanol maser data is taken
  from \citet{rygl09}.
\label{fig-demag}
}
\end{figure}

The large number of Zeeman pairs identified in transitions of OH
allows a systemic maser velocity map of ON~1 to be constructed, as
shown in Figure~\ref{fig-demag}.  Hydroxyl maser velocities have been
corrected for Zeeman splitting, which can be large (several
km\,s$^{-1}$ between LCP and RCP components) in the ground state.
Methanol maser positions from \citet{rygl09} and \citet{xu09} are
included.  No such correction is required for methanol, which is a
nonparamagnetic molecule.  There is a trend for LSR velocities to
increase toward the west along the southern line of maser spots,
although not all Zeeman pairs and methanol masers conform to this
pattern.

\subsubsection{Alignment of Maser Positions}

In aligning the 6.0~GHz masers with the \citet{fishreid07} 1.6~GHz
masers, we note that there is no particular reason to believe that the
position of the brightest 6035~MHz will be precisely coincident with
the location of the brightest 1665~MHz maser to submilliarcsecond
accuracy.  The apparent centroids of Zeeman components may appear at
slightly different locations.  For instance, the separation between
the components of 6.0~GHz Zeeman pairs listed in
Table~\ref{zeeman-table} is typically a few tenths of a
milliarcsecond.  This effect is even larger at 1665 and 1667~MHz,
where the separation between components of a Zeeman pair can be a few
milliarcseconds.  Assuming that there is no large-scale organization
that systemically shifts LCP spots in one direction and RCP spots in
the opposite direction, we can define a best statistical alignment of
the 6.0 and 1.6~GHz maser frames by finding the single vector (right
ascension and declination) offset to add to the positions of one of
the frames to minimize the sum of the squares of the 6.0~GHz--1.6~GHz
position differences among the Zeeman associations.  We use the data
in Table~\ref{zeeman-table}, using the average of RCP and LCP
positions within one frequency as the central location of the maser
emission at that frequency in each Zeeman association.  When only one
sense of circular polarization is detected in a Zeeman association in
a frequency, we use that position as the assumed central location of
the maser emission at that frequency.  We find that the two frames are
best aligned when the 1665~MHz reference feature of \citet{fishreid07}
is assumed to be 1.7~mas north of the 6035~MHz reference feature.  In
this frame, the median 1.6/6.0~GHz offset in the Zeeman associations
is approximately 2.2~mas.

The offsets between 1.6 and 6.0~GHz Zeeman associations are consistent
in magnitude with proper motions of masers in ON~1.  The
\citet{fishreid07} data were observed 3.74~yr prior to the epoch of
observations of the 6.0~GHz masers reported here.  A 1.7~mas apparent
offset between the epochs could correspond to a motion of
5.5~km\,s$^{-1}$ at a distance of 2.57~kpc \citep{rygl09}.  This is
comparable to the 1.6~GHz maser motions seen in \citet{fishreid07}.
Although the lack of phase-referencing in the ground-state
observations precludes rigorous analysis of the offsets in terms of
proper motions, the general pattern is confirmed by the 6.7~GHz
methanol observations of \citet{rygl09}, which show that the relative
speed of the northern and southern methanol masers in the plane of the
sky is $9.4$~km\,s$^{-1}$.

\subsection{Comparison with W3(OH)}

It is instructive to compare ON~1 with W3(OH), the other massive
star-forming region hosting a large number of 6.0~GHz masers observed
with a modern VLBI array.  There are numerous similarities between the
properties of 6.0~GHz masers in ON~1 and W3(OH), some of which have
already been commented on in \S\S~\ref{results} and \ref{discussion}.
For instance, 6030~MHz masers in ON~1 are rare and are only found in
the presence of 6035~MHz masers.  This is consistent with observations
of W3(OH), which found that 6030~MHz masers are nearly always found in
close spatial association with 6035~MHz masers and imply similar
physical conditions \citep{fishsjouwerman07}.

However, one important contrast between the OH masers in ON~1 and
W3(OH) is that in the former, all 6035~MHz masers are found in close
association with 1665~MHz masers.  This stands in stark contrast with
W3(OH), where many 6035~MHz maser spots are found without accompanying
ground-state emission, including some (particularly in the NE and SE
regions) where no ground-state masers are located within hundreds of
AU \citep{fishsjouwerman07}.  Theoretical modelling suggests that
co-propagation of 1665 and 6035~MHz masers (as observed throughout
ON~1) can occur under a wide range of conditions, but that the
presence of 6035~MHz masers without 1665~MHz masers may require a
higher density and/or higher OH column density
\citep[e.g.,][]{gray92,cragg02}.  It is therefore probable that the OH
masers in ON~1 are on the whole tracing slightly less dense material
than in W3(OH).

Collisions may also play a partial role in exciting the 6035~MHz
transition \citep{gray92}.  It is possible, for instance, that the
masers just south of the \ion{H}{2} region trace a shock front.  The
appearance of ground-state, satellite-line masers in this region
\citep{nammahachak06} may also be indicative of a shock
\citep{gray92,pavlakis96a,pavlakis96b}.

Methanol maser excitation is predicted to occur over a similar range
of densities \citep{cragg02}.  The tentative detection of a magnetic
field strength of $-18$~mG in methanol Zeeman splitting by
\citet{green07} may be indicative of a slightly higher density locally
at the methanol maser site in the southern line of masers in ON~1.
Methanol emission is also seen intermixed with OH masers in W3(OH)
\citep{harveysmith06}.

In W3(OH), 31\% of the 6.0~GHz masers are 6030~MHz masers, which are
typically weaker than their 6035~MHz maser counterparts by a factor of
several \citep{fishsjouwerman07}.  In ON~1, only 10\% of the 6.0~GHz
masers are in the 6030~MHz transition.  This may not be a significant
difference, due both to the smaller number of total maser spots in
ON~1 and the fact that the 6.0~GHz masers are weaker than in W3(OH),
meaning that a larger fraction of the presumed 6030~MHz maser
population in ON~1 may be below our detectability threshold.  The
modelling of \citet{pavlakis00} and \citet{cragg02} show that the
parameter spaces under which 6030 and 6035~MHz masers are likely to be
excited in star-forming regions are very similar.  For identical
excitation conditions (as would occur if 6030 and 6035~MHz masers were
found in spatial overlap), \citet{cragg02} find that the brightness
temperature of the 6030~MHz is typically a factor of a few smaller
than that of the 6035~MHz maser.  However, 6030~MHz masers are found
only in the west, despite the fact that all maser clusters in ON~1
(northern, central, western, and the line of masers in the south) host
6035~MHz masers that are at least a factor of 10 brighter than our
detection threshold.  It is not clear what, if any, are the
differences among the physical conditions of these clusters that
allows 6030~MHz maser formation only in the west.

\subsection{The Structure of ON~1}

The view of ON~1 has evolved with increasingly better observational
data.  \citet{israel83} suggested that ON~1 was an example of isolated
massive (spectral type B0.5) star formation.  More recent infrared
\citep{kumar02,kumar03,kumar04}, centimeter, and submillimeter
observations \citep{su04,su09} indicate that the environment of ON~1
contains numerous dust cores associated with other intermediate-mass
or high-mass star formation.  Multiple molecular outflows and
high-velocity water maser structures add further complexity to the
surroundings of the \ion{H}{2} region, suggesting that the source(s)
in the \ion{H}{2} region are gravitationally bound to other nascent
and possibly evolved stars in the region \citep{nagayama08,su09}.  The
complicated environment of ON~1 renders it difficult to interpret the
OH masers surrounding the \ion{H}{2} region.

Nevertheless, based on the OH maser distribution in multiple
transitions, it is likely that the \ion{H}{2} region is associated
with at least two distinct excitation sources.  The highly excited
13441~MHz transition of OH produces masers near both 0 and
$14$--$15$~km\,s$^{-1}$ \citep{baudry02,fish05}.  Archival VLBA
observations (experiment code BB137) of this transition were
unsuccessful in detecting the emission, so no interferometric map of
the 13441~MHz masers exists.  However, VLBI maps of 6.7~GHz methanol
masers, which show a similar double-peaked spectrum at 0 and
$14$--$15$~km\,s$^{-1}$, are consistent with emission in the northern
and southern groups (\citealt{sugiyama08,rygl09}; see also
\citealt{green07}), as shown in Figure~\ref{fig-demag}.  The existence
of a 12~mG magnetic field, large for OH masers, in the north argues
that the northern masers (Hh, Ii, and J) are located close to one of
these excitation sources.  The magnetic field measurements in the
south are less conclusive, although if the $-18$~mG magnetic field
inferred from methanol Zeeman splitting \citep{green07} is correct,
the bright 6035~MHz reference feature (K) may be located closest to
the southern excitation source.  We note the possibility, however,
that the large magnetic field strengths and existence of
highly-excited OH masers in the north and south may simply be due to
dense molecular clumps on opposite sides of an \ion{H}{2} region
containing only one excitation source.

The southern masers, which constitute the bulk of the observed OH
masers in ON~1, are probably best understood in the context of the
\citet{elitzur78} model of OH masers forming in the compressed region
between the ionization and shock fronts around an expanding \ion{H}{2}
region.  These masers appear at the edge of the \ion{H}{2} region as
observed at $\lambda = 1.3$~cm and are located not too far from the
edge of the dust cavity evacuated by the \ion{H}{2} region
\citep{su09}.  The systemic velocity of the molecular cloud is fairly
well established to be $v_\mathrm{LSR} = 11$--$12$~km\,s$^{-1}$ from a
large variety of dense molecular tracers including ammonia,
formaldehyde, CS, SiO, thermal methanol, and Class~I methanol masers
\citep[e.g.,][]{haschick89,haschick90a,haschick90b,plume92,anglada96,macleod98,kalenskii01,araya02}.
The line-of-sight velocity of the southern masers is close to the
assumed systemic dust velocity of 11--12~km\,s$^{-1}$ but slightly
redshifted (Figure~\ref{fig-demag}), with the offset from systemic
ranging from less than 1 to over 3~km\,s$^{-1}$.  This (line-of-sight)
velocity offset is consistent in magnitude with the 3.6~km\,s$^{-1}$
(transverse) proper motion inferred by \citet{rygl09} from methanol
maser observations and the expansion of the \ion{H}{2} region.
Unfortunately, the linear polarization in this region (from this work
as well as \citealt{fish05b} and \citealt{green07}) is complicated
enough to prevent definitive identification of the large-scale
magnetic field direction in the plane of the sky along the line of
southern masers.

The northern masers are associated with the largest magnetic field
strength seen in the source ($-12$~mG at spots Ii).  More so than in
other maser clumps, the northern masers divide neatly into regions
where only the LCP component is seen at 1665~MHz and regions where the
RCP component is seen.  After the reference feature, the 6035~MHz
Zeeman association with the largest component flux ratio is located in
the northern cluster.  These features may be explainable by correlated
magnetic fields and velocity gradients \citep{cook66} or by Zeeman
overlap \citep{deguchi86}.  There is also a relative dearth of linear
polarization in this region, at both 6035~MHz and 1665~MHz
\citep{fish05b}.  Indeed, the Zeeman splitting from the large magnetic
field at spot Ii is larger than the maser line width, which is
predicted by Zeeman overlap to produce masers that are very highly
circularly polarized \citep{deguchi86}.

Understanding the northern masers may require understanding both the
multiplicity of massive stars in the \ion{H}{2} region and the
velocity of the \ion{H}{2} region.  \citet{zheng85} measured the
velocity to be $+5$~km\,s$^{-1}$ at 14.7~GHz, although there is reason
to believe that this velocity may be significantly blueshifted from
the systemic velocity of the massive star(s) in the \ion{H}{2} region
due to optical depth effects in the expanding plasma \citep[see
  discussion in][]{fishreid07}, which are visible in the $\lambda =
3.6$ and 1.3~cm continuum maps of \citet{su09}.  Three possibilities
arise: the \citet{zheng85} velocity may indeed trace the velocity of a
single massive star within the \ion{H}{2} region; there may be only
one excitation source in the \ion{H}{2} region, but its velocity is
not given by the 14.7~GHz recombination line; or there may be more
than one excitation source within the \ion{H}{2} region.

In the first case, if there is only one source of excitation at
$v_\mathrm{LSR} \approx +5$~km\,s$^{-1}$, the northern and southern
masers are presumably both associated with it.  In this case, the
interpretation of \citet{fishreid07} is probably correct: maser
motions are dominated by the expansion of the ionized region and an
additional rotational component, each having a magnitude of several
km\,s$^{-1}$.  As is usual in massive star-forming regions, not all OH
maser velocities fit nicely into the large-scale velocity pattern, but
it is plausible that the differences can be explained by projection
effects and inhomogeneities in the density profile of the molecular
material.

In the second case, the systemic velocity of the ON~1 system, given by
the presumed lone massive star inside the \ion{H}{2} region, is likely
close to the systemic molecular velocity of $11-12$~km\,s$^{-1}$.
Recombination line studies of other ultracompact \ion{H}{2} regions
show that centimeter-wavelength recombination lines are typically
blueshifted compared to millimeter-wavelength recombination lines
\citep[see discussion in section 3.3 of][]{sewilo08}.  Since the
optical depth of the \ion{H}{2} region decreases with frequency,
higher-frequency recombination lines are sensitive to the center of
the \ion{H}{2} region (where the massive star is likely located),
while lower-frequency lines trace the expanding outer portions of the
\ion{H}{2} region and are therefore blueshifted
\citep{berulis83,welch87,keto95}.  If the same effect operates in
ON~1, the velocities of the northern masers (mostly near
0~km\,s$^{-1}$) differ by more than 10~km\,s$^{-1}$ from the rest of
the OH masers, the systemic velocity of the molecular material in the
vicinity, and presumably the velocity of the star in the center of the
\ion{H}{2} region.  It is possible, then, that the northern masers
trace an outflow, possibly with a helical magnetic field.
Ground-state RCP and LCP features have a tendency to segregate on
opposite sides of the presumed structure, which could be explained if
the magnetic field wraps around the outflow.  Unfortunately the
magnetic field direction and strength at spot Hh are poorly determined
due to the low signal-to-noise ratio of the LCP spot, so it is not
possible to confirm that the polarity of the magnetic field reverses
across the northern structure.

In the third case, there are at least two massive stars in the
\ion{H}{2} region, each with associated OH masers.  The
\citet{zheng85} recombination line velocity may represent an average
of the velocities of the two sources, although optical depth effects
may still be important.  Based on both the recombination line velocity
and the fact that the northern masers (unlike the southern masers) are
projected atop the \ion{H}{2} region, it is probable that the northern
excitation source is in front of the southern excitation source.
Either the expansion/rotation or outflow interpretation of the maser
distributions may apply, depending on the relative velocities of the
two excitation sources.

In contrast with observations of W3(OH), the 6.0~GHz masers in ON~1 do
not clearly trace additional structures that were not visible at
1.6~GHz, limiting their ability to provide additional information on
the context of the various maser clusters in the source.  Further
understanding of how the maser clusters are associated with one
another will require higher-resolution observations of the nonmasing
material.  Sensitive observations of high-frequency recombination
lines in the \ion{H}{2} region may be illuminating regarding the
internal structure of the ionized region, especially if done with
enough spatial resolution to understand the internal dynamics of the
ionized region.  ALMA observations of dense molecular tracers may also
assist in providing the context in which star formation is taking
place in ON~1.

\section{Conclusions}

We have imaged the 6030 and 6035~MHz OH masers in ON~1 at VLBI
resolution.  The distribution of these excited-state masers is similar
to that of the ground-state masers.  Unlike in the similar massive
star-forming region W3(OH), 6.0~GHz masers are not found in the
absence of 1665~MHz masers, perhaps suggesting that ON~1 does not have
analogues of the highest-density knots found in W3(OH).  The 6.0~GHz
masers are spatially coincident with 1665~MHz masers to within a few
milliarcseconds.  Magnetic fields strengths obtained from Zeeman
splitting at 6.0 and 1.6~GHz are usually consistent to better than
1~mG, suggesting that multitransition Zeeman associations are
acceptable for obtaining estimates of the local magnetic field.

Our observations confirm the existence of a strong ($-12$~mG) magnetic
field among the northern, blueshifted masers, as suspected from early
EVLA observations of ON~1 \citep{fish07}.  The large magnetic field
here, as well as the distribution of methanol maser spots and presumed
distribution of the highly-excited 13\,441~MHz OH masers, may indicate
that the northern and southern masers are excited by two different
sources within the \ion{H}{2} region or that the northern masers trace
and outflow.  However, other scenarios, such as large-scale expansion
and rotation of all groups of masers in ON~1, cannot be ruled out from
the data at hand.  The overall structure of ON~1 remains uncertain in
the absence of sensitive, high-resolution observations of the
nonmasing material.

\acknowledgments

The National Radio Astronomy Observatory is a facility of the National
Science Foundation operated under cooperative agreement by Associated
Universities, Inc.  The European VLBI Network is a joint facility of
European, Chinese, South African and other radio astronomy institutes
funded by their national research councils.

{\it Facilities: \facility{EVN}, \facility{GBT}, \facility{EVLA}}

\end{document}